\documentclass{article}

\usepackage{arxiv}

\usepackage[utf8]{inputenc} 
\usepackage[T1]{fontenc}    
\usepackage{hyperref}       
\usepackage{url}            
\usepackage{booktabs}       
\usepackage{amsfonts}       
\usepackage{nicefrac}       
\usepackage{microtype}      
\usepackage{lipsum}
\usepackage{graphicx}
\usepackage{tabularx}
\usepackage{amsmath}
\usepackage{float}
\usepackage{subcaption}
\graphicspath{ {./images/} }

\title{Student-LLM Interactions During Math Learning: Limited Epistemic Proactivity Overall, but Within-chat Change Predicts Performance Without AI}

\author{
 Rania Abdelghani \\
  Hector Research Institute of Education Sciences and Psychology\\
  University of Tübingen\\
  Germany \\
  \texttt{rania.abdelghani@uni-tuebingen.de} \\
   \And
 Peter Kaiser \\
  Hector Research Institute of Education Sciences and Psychology\\
  University of Tübingen\\
  Germany \\
  \texttt{pe.kaiser@uni-tuebingen.de} \\
  \And
 Kou Murayama \\
  Hector Research Institute of Education Sciences and Psychology\\
  University of Tübingen\\
  Germany \\
  \texttt{k.murayama@uni-tuebingen.de} \\
}

\begin{document}
\maketitle
\begin{abstract}
Large language models (LLMs) are increasingly used for learning, raising a key question: can learners remain \emph{epistemically proactive}, retaining responsibility for the cognitive and regulatory activities through which learning is produced while working with AI? Because this responsibility may change as interactions unfold, we examine epistemic proactivity as a developing process. We conceptualize epistemic proactivity through self-regulated learning (SRL) moves, complemented by help-seeking strategies and manifestations of epistemic vigilance. We analyzed these behaviors in 880 substantive turns from 112 secondary-school students completing open-ended mathematical-modeling practice with a general-purpose LLM (Mistral). Overall, students showed limited observable epistemic proactivity: their interactions were strongly request-dominated, with planning, monitoring, and evaluation relatively rare, although low-quality AI responses elicited more explicit evaluation. Temporal analyses qualified this picture: explicit SRL-related moves increased modestly over the chat, but help-seeking strategies became less understanding-oriented. Crucially, these temporal trajectories in help-seeking strategies predicted their post-test performance without AI beyond pre-test and aggregate patterns: students whose requests became more understanding-oriented performed better once AI assistance was removed. Thus, while students had a limited epistemic proactivity overall, its temporal development still was a critical predictor of post-test performance without AI. These results underscore the importance of examining the temporal organization of epistemic proactivity, rather than isolated prompt types, for understanding AI-supported learning. We release our fully-anonymized dataset publicly in \href{https://osf.io/p4y8c/overview?view_only=756a198d244c4ad2a25a9866359079c0}{this anonymous OSF repository}.
\end{abstract}


\section{Introduction}
Students are increasingly using general-purpose large language models (LLMs) for learning~\cite{maheux2026generative}. While they can use LLMs to improve immediate task performance, their use does not necessarily support the knowledge-construction processes needed for durable learning---such as planning, reasoning, and structuring problems~\cite{Yan2025DistinguishingAI}. Learners may instead delegate these learning-relevant cognitive and regulatory work to the tool without an immediate or visible cost to task completion~\cite{yan2026agentivism}. A central question for AI-supported learning is therefore whether learners can retain responsibility for these cognitive and regulatory processes through which learning occurs, while benefiting from the assistance these systems provide.

This concern is increasingly discussed in terms of learner agency: whether learners remain able to direct and take responsibility for their own learning when working with AI~\cite{yan2026agentivism}. Agency, however, is a broad and context-dependent construct that has been conceptualized as encompassing regulatory and reflective abilities, agentic actions such as selective and responsible engagement, and motivational characteristics such as self-efficacy and persistence~\cite{xia2025developing}. This breadth makes agency an important overarching concern, but difficult to investigate empirically. This challenge is even more pronounced in complex learning environments such as GenAI-supported ones where the system itself can reshape several components of learners' agency~\cite{fan2025beware}.

We therefore focus on one specific epistemic dimension of learner agency, which we term \emph{epistemic proactivity}: the extent to which learners retain responsibility for initiating, monitoring, advancing, and evaluating their own knowledge-building processes with AI~\cite{darvishi2024impact,abdelghani2026illusion}. Epistemic proactivity therefore distinguishes initiating a learning trajectory from primarily reacting within one proposed by the system~\cite{gerlich2025ai}. In doing so, it also distinguishes behavioral from epistemic engagement during AI-supported learning: a learner may remain behaviorally agentic---asking questions, evaluating suggestions---while becoming epistemically reactive if the AI is increasingly determining what should be explored next, and how the interaction should proceed~\cite{gerlich2025offloading,lodge2026artificial}. We examine this dimension through self-regulated learning (SRL) as the overarching regulatory framework, complemented by help-seeking strategies and epistemic vigilance as finer-grained accounts of what learners ask AI to contribute and how they evaluate its contributions.

Importantly, epistemic proactivity is unlikely to remain fixed during a learning process: it may shift as task demands, learners' resources, and experience with the tool evolve~\cite{ruggeri2022introduction}. Studying it therefore means asking how it develops across an interaction, not only how often particular behaviors occur. However, such examinations are still missing in literature; and, specifically, two main gaps remain. First, process-level analysis of student--AI interactions is still limited for secondary-school students compared with university populations, despite their growing use of these systems~\cite{maheux2026generative,misiejuk2026cognitive}. Second, little is known about whether such interaction dynamics relate to learning beyond the performance achieved while AI assistance is available. This question is especially relevant during adolescence, when capacities required for an epistemically-active use of AI are still developing~\cite{oakley2026memory}. Against this background, we studied 112 Grade-9 students using a general-purpose LLM during an open-ended mathematical-modeling activity between a pre-test and post-test completed without AI. We address two questions: to what extent do students remain epistemically proactive while learning with AI, and whether the way epistemic proactivity develops over time is associated with what they can subsequently do without AI. We operationalize epistemic proactivity in conversational traces through SRL, complemented by finer-grained analyses of help-seeking strategies and epistemic vigilance. By answering these questions, we aim to clarify when AI-supported interaction reflects productive learning rather than successful performance with assistance, and to inform the design of scaffolds for adolescent learners.

\section{Related Work}
\label{sec:related-work}
\subsection{Generative AI in Mathematics Education: Between Affordance and Reliance}
Mathematics makes the tension between the affordances and risks of LLM-supported learning particularly visible. LLMs can provide explanations, worked examples, and feedback across mathematical problem-solving activities~\cite{pepin2025scoping,dilling2024using,gallart2026modelling}. Yet mathematical learning requires more than obtaining correct solutions: learners must learn how to structure problems, select and justify representations, interpret results, and evaluate whether they meaningfully address the problem~\cite{schoenfeld2007mathematical}. This is especially salient in mathematical modeling for instance, a core mathematical competency where learners must progressively structure an initially open situation rather than follow a predefined procedure~\cite{blum2009mathematical}. While LLMs can support these processes, they can also perform them on the learner's behalf, when their assistance is not calibrated to the learner's current knowledge and background~\cite{bulut2026ai}.Their outputs may also be inaccurate, oversimplified, or pedagogically weak, creating a further risk when such problems go undetected~\cite{pepin2025scoping,schorcht2026no}.

Empirical work illustrates this tension. Pre-service mathematics teachers have been found to misjudge erroneous ChatGPT responses and subsequently produce incorrect solutions~\cite{noster2024pre}. Among secondary students solving mathematical-modeling problems with ChatGPT, AI use concentrated on early phases such as simplifying and structuring, whereas interpretation and validation were comparatively rare, potentially suggesting that AI use led them to bypass such important processes~\cite{bulut2026ai}. Furthermore, a large-scale study by Bastani et al. found that unrestricted GPT-4 improved performance while assistance was available, but this advantage did not persist when students later worked without AI, whereas pedagogical guardrails largely mitigated the effect~\cite{bastani2025generative}. These results suggest that students delegated the full epistemic work to the tool when they had access to it. Related work beyond mathematics similarly suggests that delegating learning-relevant cognitive work to LLMs can reduce learners' processing and integration of information~\cite{stadler2024cognitive,darvishi2024impact,melumad2025experimental}.

Together, these findings suggest that AI can facilitate immediate performance while shifting learning-relevant work from learner to system. Compared with classical forms of cognitive offloading such as externalizing memory or information retrieval~\cite{sparrow2011google}, LLMs broaden what can be delegated to include problem structuring, reasoning, information synthesis, and decisions about how an inquiry should proceed~\cite{gerlich2025offloading,melumad2025experimental}. Productive AI-supported learning therefore cannot be understood from task success alone: it also requires examining how epistemic responsibility for initiating, advancing, and evaluating learning--i.e. epistemic proactivity-- is shared between learner and AI.

\subsection{Epistemic Proactivity through Self-Regulation: Help-Seeking and Epistemic Vigilance}
We use self-regulated learning (SRL) as the overarching framework for understanding how epistemic proactivity unfolds during AI-supported learning. SRL concerns how learners set goals, plan how to pursue them, monitor their strategies and understanding, and evaluate progress to guide subsequent learning~\cite{zimmerman2002becoming,Zimmerman2008InvestigatingProspects}. These processes closely correspond to the forms of epistemic responsibility captured by epistemic proactivity. Effective self-regulation is an important topic in education research as it is generally associated with stronger learning and engagement~\cite{futterer2026enhancing}. In student--AI dialogue, explicit expressions of plans, uncertainty, requests, and evaluations can therefore provide observable evidence of regulation, although regulation that remains internal cannot be directly observed. We therefore focus our analysis on detecting observable expressions of planning, monitoring, requesting information, and evaluation in student--AI dialogue. 

Within this broader regulatory process, \emph{help-seeking} provides a finer-grained account of \textit{how} learners implement their learning strategies, once they identify a learning goal~\cite{nelson1986help}. The help-seeking lens is particularly informative for epistemic proactivity because different types of requests leave different amounts and types of epistemic work with the learner. For example, help-seeking research distinguishes instrumental assistance, which supports continued independent reasoning, from executive assistance, which transfers more of the task to the external source~\cite{nelson1986help}. Question-asking research similarly distinguishes requests supporting explanation and knowledge construction from those primarily seeking convergent information~\cite{graesser1994question,alaimi2020pedagogical}. In this work, we therefore distinguish conceptual and procedural requests oriented toward advancing understanding from answer- and verification-seeking requests oriented more toward epistemic closure~\cite{graesser1994question}. This perspective also avoids equating prompt sophistication with productive learning: an elaborate prompt may delegate substantial reasoning, whereas a brief request may target a precise conceptual gap~\cite{denny2024generative}.

A second specific lens within this regulatory process is \emph{epistemic vigilance}: evaluating communicated information before accepting or integrating it~\cite{sperber2010epistemic}. Such appraisal can regulate subsequent activity, for example by prompting learners to seek clarification, reformulate a question, or test an alternative~\cite{ronfard2018question}. In AI-supported learning, epistemic vigilance is therefore relevant not only to whether learners detect problematic assistance, but also to whether this appraisal informs how they subsequently regulate the interaction. Epistemic vigilance may depend on domain knowledge, the quality of the information received, and perceived source reliability~\cite{kidd2023}. Accordingly, in this work, we aim to examine conversational traces of both evaluative detection and subsequent regulatory response.

In short, our framework is hierarchical: SRL captures the broader regulatory functions through which learners plan, monitor, seek information, and evaluate their learning. Within this process, help-seeking further characterizes \emph{what} learners ask the AI to contribute when seeking assistance, while epistemic vigilance characterizes \emph{how} they evaluate and respond to what the AI provides. Examining these levels together allows us to characterize epistemic proactivity without assuming that any single behavior or prompt is inherently productive: what matters is how learners retain the overall responsibility for identifying what needs to be understood, deciding what to delegate, and using AI contributions to regulate subsequent learning.

\subsection{Epistemic Proactivity as an Evolving Process in Student--AI Dialogue}
Epistemic proactivity is unlikely to remain fixed throughout an interaction. As learners gain experience with an AI, and as task demands and available cognitive resources change, they may adapt what they regulate themselves, what assistance they seek, and how carefully they evaluate the system's contributions~\cite{misiejuk2026cognitive}. They may increasingly delegate structuring or reasoning when the system appears capable of performing these activities, or reclaim responsibility when its assistance proves insufficient or unsatisfactory~\cite{gerlich2025offloading,Joshi2025ELI-Why:Explanations}.

Importantly, such delegation is not inherently detrimental. Cognitive offloading can reduce demands that would otherwise exceed learners' available resources~\cite{risko2016cognitive}, potentially freeing effort for other learning-relevant activity~\cite{Seufert2018TheLoad,ruggeri2022introduction}. This is also in line with the broader "assistance dilemma": external support can be beneficial for learning if it is carefully designed to enable progress without replacing the learner activity required for learning~\cite{koedinger2007exploring}. With LLMs, the relevant question is therefore not simply whether learners delegate cognitive or regulatory work, but \emph{what} they delegate, \emph{when}, and whether epistemic responsibility is subsequently retained or reclaimed.

This makes the analysis of temporal organization of AI-supported learning important. Recent studies have begun to move beyond behavioral frequencies using process-oriented methods that capture how interaction patterns unfold. For instance, work in~\cite{alamray2026exploring} coded student--GenAI exchanges using Bloom's taxonomy and combined Epistemic Network Analysis with frequency-based Transition Network Analysis to examine both co-occurring cognitive processes and their sequential transitions, revealing recurrent rather than strictly linear patterns of cognitive engagement. Similarly, work in~\cite{misiejuk2026cognitive} followed students across four assignments, qualitatively coded their prompts, and used Co-Occurrence Network Analysis to examine how prompting and cognitive-offloading patterns developed with repeated AI use. Related work has used sequential pattern mining and transition analysis to trace how epistemic functions are organized across student--AI dialogue~\cite{zhou2026catalyst}. Complementing these observational approaches, experimental evidence also suggests that ordering matters: learners who generated their own ideas before consulting ChatGPT subsequently performed better without AI than those given unrestricted access from the beginning~\cite{wong2026think}. Together, these studies show that AI-supported learning can be examined at multiple temporal scales, from local transitions between interaction states to broader changes across repeated use and experimentally structured sequences. Such analyses can reveal patterns of AI use and cognitive engagement that are obscured by overall behavioral frequencies, particularly as students' prompting and offloading strategies change with repeated experience.

While informative, a key question remains under-explored in this literature: whether such changes during AI use are associated with what learners can subsequently do without AI. This is important as successful assisted performance does not establish that knowledge or skills remain available once assistance is removed~\cite{Yan2025DistinguishingAI,bastani2025generative}. Existing temporal work provides limited evidence on whether/what features student-AI regulation carry such information beyond learners' initial domain knowledge and what can already be inferred from aggregate patterns of AI use.

We therefore examine both broader within-chat changes and local sequential organization, and test whether these dynamic features are associated with later performance without AI beyond pre-test performance and static interaction patterns. As discussed above, we focus on adolescent school students because this developmental period may be particularly sensitive to over-delegation to AI: the very same cognitive and regulatory capacities needed to decide what to delegate, monitor AI assistance, and retain control over learning are themselves still developing~\cite{oakley2026memory}.

\section{Current Study and Research Questions}
We examine epistemic proactivity among Grade-9 school students using a general-purpose LLM during open-ended mathematical modeling practice situated between a pre-test and post-test without AI. Mathematical modeling provides a particularly suitable context because learners must not only apply mathematical knowledge, but also structure the problem, select and revise strategies, interpret results, and evaluate their validity~\cite{kaiser2006global,blum2009mathematical,ferri2010mathematical}. Moreover, rather than being asked to use the LLM to solve a specific problem, students were instructed to use the interaction to advance their understanding of mathematical modeling in preparation for an exam. Decisions about what needed to be learned, how to organize the interaction, and what assistance to seek were therefore largely left to the students. Together, these two design choices allow us to examine students' spontaneous expression of epistemic proactivity, as no particular interaction strategy, specific problem-solving goal, or pattern of assistance-seeking was imposed.

Drawing on the framework developed above, we examine epistemic proactivity through students' SRL functions and the epistemic function of their help-seeking requests, complemented by an exploratory analysis of how students respond to low-quality AI assistance as an indicator of their epistemic vigilance. Beyond these overall patterns, we examine two complementary scales: the local sequential organization of behavior and broader within-session changes in SRL and HS. Finally, we test whether these dynamic features account for subsequent post-test performance without AI, beyond pre-test performance and static patterns of AI use.

Accordingly, we address three research questions:

\begin{enumerate}
\item \textbf{RQ1: Epistemic Proactivity Profile.} What is overall profile of SRL, Help-seeking, and epistemic vigilance of high-school students' interactions with the LLM?

\item \textbf{RQ2: Within-Session Dynamics of Epistemic Proactivity.} How are high-school students' SRL and help-seeking behaviors organized locally and how do they change over the course of the interaction?

\item \textbf{RQ3: Post-Interaction Performance without AI assistance and Links with Epistemic Proactivity.} What static and dynamic features of high-school students' epistemic proactivity during AI use are associated with their subsequent post-test performance without the help of AI beyond, pre-test performance?
\end{enumerate}

\section{Methods}
\subsection{Participants}
Participants were 112 Grade-9 high-school students from three public Gymnasium schools (academically oriented secondary schools) in Baden-Württemberg, Germany. Students were 14--15 years old ($M=14.67$). For analyses involving post-test performance, we only had a sample of 95 students as some them had missing data for either one of the tests. Students were familiar with the underlying mathematical concepts from their regular curriculum but had not previously worked on the specific modeling skills used in the study.

The study was approved by the Ethics Committee of the University of Tübingen (AZ: A2.5.4-438{\_}hb) and by the Ministry of Culture, Youth and Sports of Baden-Württemberg. Written informed consent was obtained from all students and guardians prior to participation. 

\subsection{Study Material and Procedure}
\label{sec:procedure}
\begin{figure}
    \centering
    \includegraphics[width=0.7\linewidth,clip]{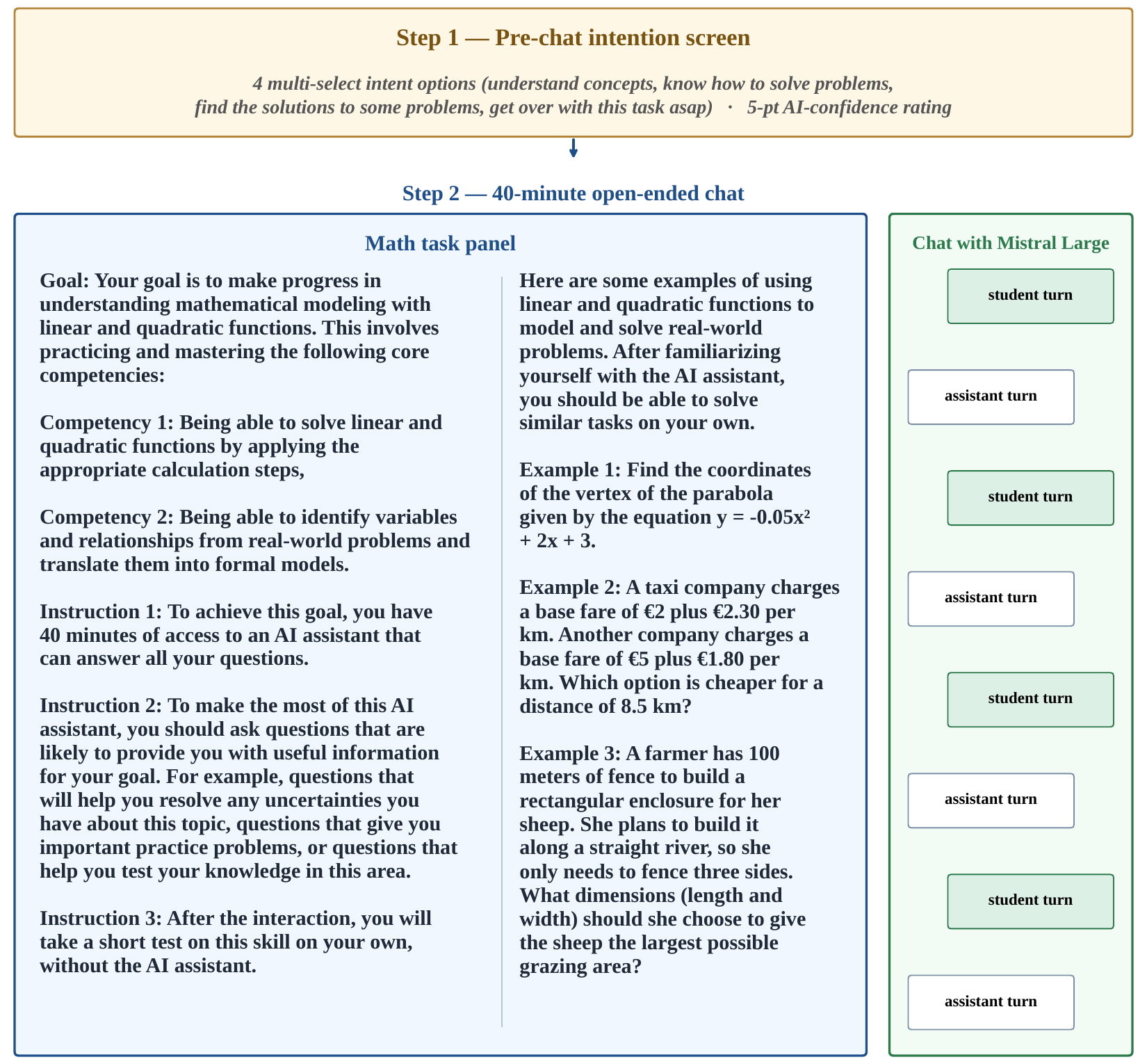}
    \caption{Overview of the study procedure, with a detailed view of the open-ended AI-supported phase.}
    \label{fig:Phases}
\end{figure}

The study was conducted during regular mathematics lessons on tablets and lasted approximately 60 minutes. Students first completed a 10-minute mathematics pre-test assessing the targeted modeling competencies. They were then presented with a screen describing the learning activity~\autoref{fig:Phases}. 

As seen in the figure, the task was explicitly framed as a \emph{learning activity}: students were told that their goal was to improve their understanding of mathematical modeling with linear and quadratic functions, rather than to solve a predefined set of problems. These topics formed part of their current mathematics curriculum and had been studied throughout the school year; they would also remain relevant in subsequent years. The activity was therefore intended to consolidate and extend competencies that students had already begun to develop. Students were instructed to identify uncertainties in their understanding of this topic, ask the tutor for assistance, and use the interaction to prepare themselves to solve mathematical-modeling problems in a subsequent post-test where AI assistance will be removed. To make this goal concrete, the instructions provided examples of questions they could ask the tutor and of the types of problems they should be able to solve on their own after they finish the training.

After reading these instructions and receiving additional explanations from the researcher, students then engaged in approximately 40 minutes of open-ended mathematical-modeling practice with the AI tutor.

No fixed sequence of problems or questions was imposed. Students were therefore responsible for deciding which aspects of mathematical modeling to explore, what assistance to request, and how to use the tutor. To ensure a minimum amount of interaction, they were required to contribute at least six task-related turns before ending the chat. Empty, nonsensical, and off-task messages were automatically filtered, excluded from this minimum, and not added to the conversation history.

Finally, students completed a 10-minute post-test without access to the AI tutor. Pre- and post-test items differed but assessed the same mathematical-modeling competencies and were selected for comparable difficulty. The items were drawn from official curriculum materials and required similar mathematical reasoning steps and application of linear and quadratic functions while differing in wording and problem context. Further details about the assessment items are provided in~\autoref{sec:questionnaires}.

\subsection{Measures}
\paragraph{Mathematical Modeling Performance}
\label{sec:questionnaires}
Students' mathematical modeling performance was assessed using pre- and post-test items drawn from the mathematics item pool developed by the German Institute for Educational Quality Improvement (IQB): \url{https://www.iqb.hu-berlin.de/de/schule/aufgaben/seki/vera-8-mathematik/}. The selected items specifically targeted the mathematical-modeling competencies practiced during the AI-supported activity. Items were selected based on the official IQB classifications of competence area, content focus, and difficulty. Different but matched items were administered at pre- and post-test to minimize repetition and practice effects while targeting comparable modeling competencies. Pre-test performance served as a baseline measure of students' modeling performance and was included as a covariate in the analyses predicting subsequent post-test performance. Responses were scored manually by the research team.

\subsubsection{Codebook Development, Validation, and Application to Student-AI Conversations}
\label{sec:codebook-dev}
\begin{figure}
    \centering
    \subfloat[\centering \textbf{Codebook Dimensions and Turn-Level Coding Scheme} ]{{\includegraphics[width=8cm]{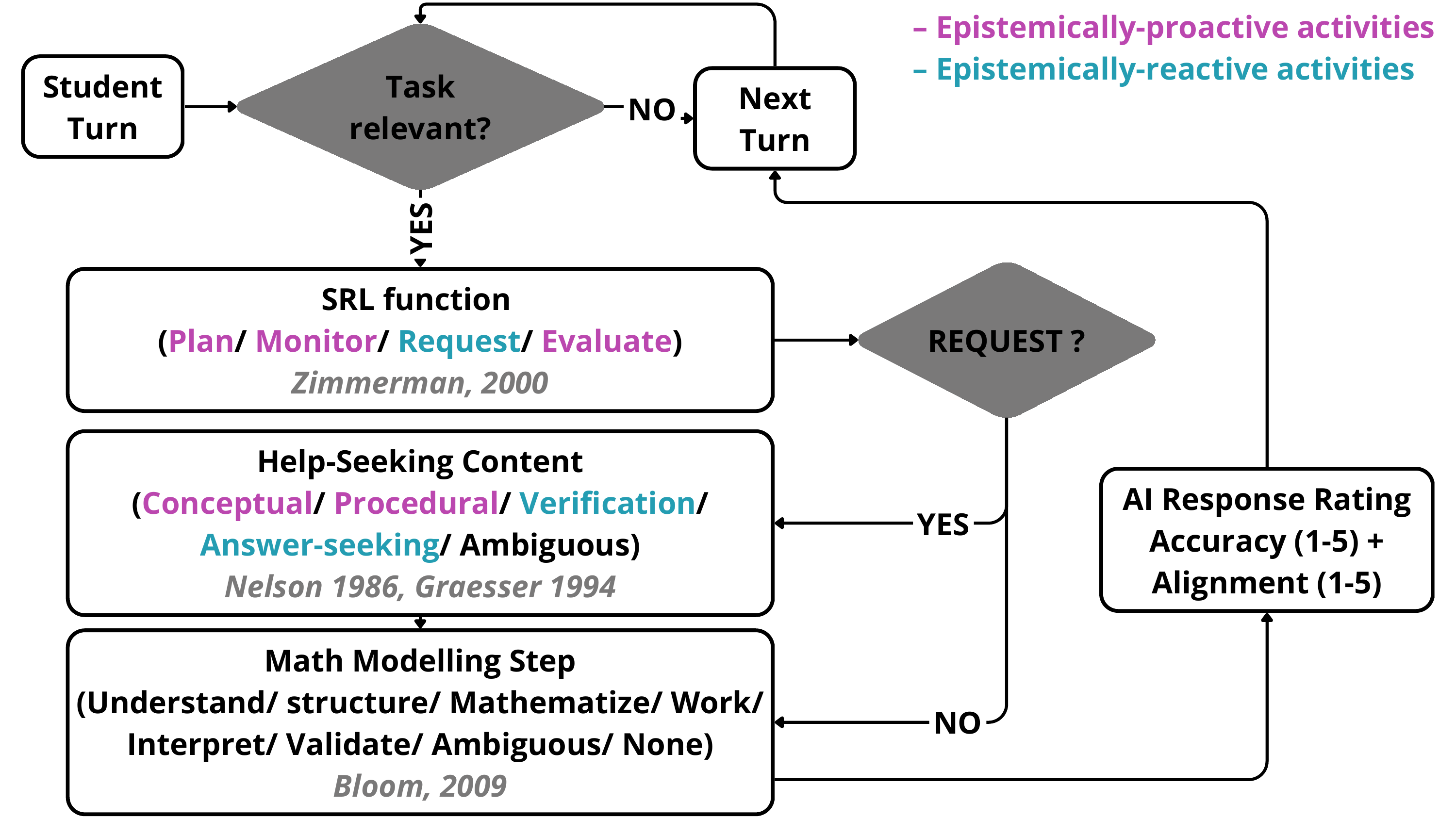} }}%
    \subfloat[\centering \textbf{Codebook development, validation, and application pipeline}]{{ \includegraphics[width=8cm]{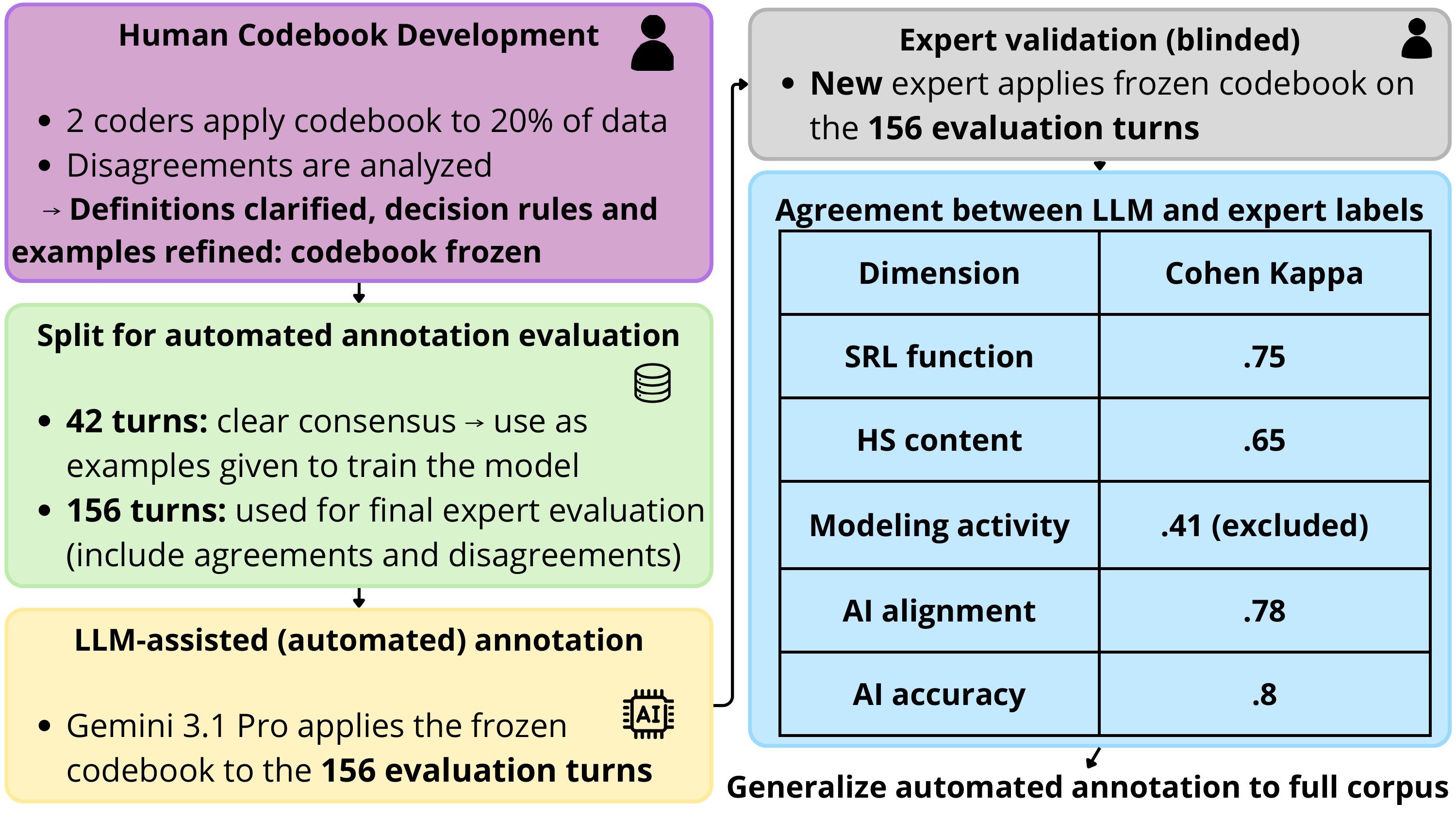} }}%
    \caption{\centering Overview of the conversational coding framework and its validation}%
    \label{fig:Coding}%
\end{figure}

\paragraph{Codebook Development.} Building on the hierarchical framework developed above, we constructed a theory-driven, turn-level coding scheme to operationalize epistemic proactivity in student--AI conversations. As summarized in sub-figure (a) of~\autoref{fig:Coding}, each student turn was first coded for its broad SRL function; SRL turns that contained assistance-seeking were then further characterized through help-seeking content. Tutor responses were separately rated for quality, and students' subsequent activity was used as an exploratory indicator of epistemic vigilance---that is, which SRL and/or help-seeking behavior followed an identified low-quality response.

\emph{SRL function} captured the regulatory role of each student turn: \textit{PLAN} for explicitly communicating epistemic goals or strategies, \textit{MONITOR} for expressing uncertainty, lack of (or achievement of) understanding, or knowledge gaps, \textit{REQUEST} for seeking assistance, and \textit{EVALUATE} for judging the correctness, relevance, or usefulness of the tutor's response. \textit{PLAN}, \textit{MONITOR}, and \textit{EVALUATE} therefore provided observable evidence of explicit learner regulation, whereas \textit{REQUEST} captured assistance-seeking. Importantly, \textit{REQUEST} was not itself treated as epistemically reactive; its epistemic function depended on what assistance was requested and how it was situated within the interaction and learning process.

Turns classified as \textit{REQUEST} were therefore further coded for \emph{help-seeking content}: conceptual, procedural, verification-seeking, or answer-seeking~\cite{nelson1986help,graesser1994question,tawfik2020role}. Conceptual and procedural requests were grouped as \emph{understanding-oriented}, whereas answer- and verification-seeking requests were grouped as more \emph{closure-oriented}. These categories do not imply that individual requests are inherently productive or unproductive; rather, they distinguish requests that tend to extend reasoning from those primarily seeking or confirming an outcome. These contrasts were subsequently used to examine how students' help-seeking changed across the interaction.

Finally, and as discussed in~\autoref{sec:related-work}, the presence of an \textit{EVALUATE} turn alone does not fully capture epistemic vigilance, which also concerns whether learners use their appraisal of information to regulate subsequent learning behavior~\cite{ronfard2018question}. Tutor responses were therefore rated on five-point scales for \emph{accuracy} and \emph{need alignment}. We then examined whether students detected inaccurate or poorly aligned assistance and how their subsequent SRL and/ or help-seeking behavior manifested in response.

\footnote{As a side note: mathematical modeling dimension based on Blum's framework~\cite{blum2009mathematical} was also initially included, but was excluded from subsequent analyses after failing to reach satisfactory validation.}

\paragraph{Codebook Validation.}
The corpus contained 988 student turns from 112 students, of which 880 substantive, task-related turns were retained for interaction analyses. Our validation procedure was designed to separate two objectives: first, developing a sufficiently clear human coding framework, and second, evaluating whether an LLM could reliably apply that finalized framework at scale.

Two trained coders initially applied a preliminary version of the codebook described above to a stratified sample of 198 turns (approximately 20\% of the corpus); see see sub-figure (a) of~\autoref{fig:Coding}. Initial inter-rater reliability was limited for several dimensions---e.g. several disagreements were found between \textit{PLAN} versus \textit{REQUEST}, \textit{MONITOR} versus \textit{REQUEST}, and among Help-Seeking categories. These disagreements were therefore used diagnostically to identify ambiguous definitions and decision boundaries in the codebook, refine category definitions, introduce ordered decision rules, and clarify prototypical and borderline cases. The codebook was then frozen.

To configure the automated annotation procedure, we selected 42 cases showing clear agreement between the two coders as in-context examples. Gemini 3.1 Pro was then prompted with the finalized codebook, ordered decision rules, rating scales, the 42 human-coded examples with clear agreement, and constrained output labels.

The remaining 156 turns were withheld from development of the automated annotator and used for the later final validation. Because the original human labels had been produced using the preliminary codebook, a third expert independently re-coded these turns using the final codebook while remaining blinded to both the initial coder labels and the LLM predictions. These annotations constituted the final expert reference. This design allowed us to evaluate the automated procedure on cases it had not seen during prompt development, while ensuring that the reference labels reflected the finalized codebook; see sub-figure (b) of~\autoref{fig:Coding}.

Agreement between the LLM and expert reference labels was assessed separately for each dimension using Cohen's $\kappa$: $\kappa=.75$ for SRL function, $\kappa=.65$ for Help-Seeking content, $\kappa=.78$ for tutor-response accuracy, and $\kappa=.80$ for need alignment. A mathematical-modeling-stage dimension based on Blum's framework~\cite{blum2009mathematical} was also initially developed, but agreement reached only $\kappa=.41$ and this dimension was therefore excluded from subsequent analyses.

Following validation, the retained automated annotation procedure was applied to all 880 substantive student turns for SRL and Help-Seeking, and to tutor responses for accuracy and need alignment. These results will be used throughout the results section. Each target turn was interpreted in the context of the full conversation. The complete codebook, prompts, model configuration, examples, and annotation scripts are provided in the supplementary materials at
\href{https://osf.io/p4y8c/overview?view_only=756a198d244c4ad2a25a9866359079c0}.

\subsubsection{Dynamic Interaction Analysis}
\label{sec:features-def}
As developed in~\autoref{sec:related-work}, epistemic proactivity concerns not only which regulatory and help-seeking processes students enact, but also how these latter are organized as the interaction unfolds. We therefore examined temporal organization at two complementary scales: broader development within the whole interaction session and local sequential organization.

\paragraph{Within-session trajectories.} To examine how students' SRL and Help-seeking developed over the interaction, each substantive student turn was assigned a normalized within-session position ranging from 0 to 1. This helps us test how the four SRL functions and four Help-seeking strategies varied across session position.

We then construct student-level trajectory indices to characterize the \emph{direction} of this change between the early and late halves of each interaction. Early turns had a normalized position $<.5$ and late turns a position $\geq.5$. For SRL, the trajectory captured whether the interaction shifted from purely \textit{REQUEST}ing information toward greater explicit regulatory activities, using \textit{PLAN}, \textit{MONITOR}, and \textit{EVALUATE}. For HS, it captured whether requests shifted from answer-seeking and verification-seeking toward more understanding-oriented conceptual and procedural assistance.

For each dimension $d$, the trajectory index was defined as:
\medskip

\noindent\makebox[\linewidth][c]{%
$\displaystyle
\Delta_d =
\left[
p_{\mathrm{late}}(O^{+}_d) - p_{\mathrm{late}}(O^{-}_d)
\right]
-
\left[
p_{\mathrm{early}}(O^{+}_d) - p_{\mathrm{early}}(O^{-}_d)
\right]
$
}

\medskip

where $O^{+}_d$ and $O^{-}_d$ denote the two theoretically contrasted orientations within each dimension. For SRL, $O^{+}$ comprised \textit{PLAN}, \textit{MONITOR}, and \textit{EVALUATE}, and $O^{-}$ comprised \textit{REQUEST}. Positive values therefore indicated a shift toward greater explicit regulatory activity, whereas negative values indicated increasing request dominance, thus suggesting potential gradual delegation of the initiation and structuring of the learning trajectory. For Help-Seeking, $O^{+}$ comprised conceptual and procedural requests (\emph{understanding-oriented}), and $O^{-}$ comprised answer-seeking and verification-seeking/ validation requests (\emph{closure-oriented}). Positive values therefore indicated increasingly understanding-oriented Help-seeking, whereas negative values indicated increasingly closure-oriented help-seeking, as students put less effort in formulating inquiries that advance learning.

These indices capture broad changes in how students distribute epistemic responsibility across the interaction. Importantly, they do not imply that individual \textit{REQUEST}, answer-seeking, or verification-seeking turns are inherently unproductive; rather, considered together, they characterize whether students increasingly retain responsibility for initiating, structuring, advancing, and evaluating their AI-supported learning. Two contrasting developments are theoretically plausible. As students become more familiar with the task and interaction, reduced demands may free cognitive resources for more effortful regulatory and understanding-oriented activity~\cite{Seufert2018TheLoad}. Conversely, as students experience the AI as capable of providing sufficient assistance, they may increasingly delegate these costly activities to the system rather than resume them themselves.

\paragraph{Local sequential organization.} Session-level trajectories capture broader directional change but not how behaviors are organized from one interaction step to the next. We therefore complement them with first-order transition analyses of SRL functions and Help-Seeking strategies~\cite{alamray2026exploring}. Observed transitions will be compared with a within-student permutation null that preserves each student's overall behavioral composition while disrupting sequential order, allowing us to identify transitions that occurred more often than expected from students' aggregate behavior alone. 

\section{Results}
\subsection{Sample Description}
The interaction corpus comprised 988 student turns from 112 students. After excluding off-task and non-substantive input (e.g., "ok thanks", "yes", etc), 880 turns were retained and coded for the interaction analyses. Descriptive statistics are reported in~\autoref{tab:measures}. 

\begin{table}
\caption{Measures taken before, during, and after the AI interaction}
\label{tab:measures}
\centering
\renewcommand{\arraystretch}{1.4}
\begin{tabular}{|>{\raggedright\arraybackslash}p{0.34\textwidth}|c|c|c|c|}
\hline
Construct & $N$ & Mean & SD & Range \\
\hline
Pre-test Math Performance (\%)        & 95  & 68.04 & 31.15 & 0--100 \\
Total Time Spent in Chat (min)        & 112 & 28.60 & 6.17  & 0--40 \\
Mean AI-response-viewing Time (s)     & 112 & 148   & 81    & -- \\
Number of Substantive Student Turns   & 112 & 7.86  & 5.84  & -- \\
Post-test Math Performance (\%)       & 95  & 53.20 & 26.40 & 0--100 \\
\hline
\end{tabular}
\end{table}

\subsection{RQ1: Interaction Profile of Epistemic Proactivity} 
\paragraph{SRL Functions and Help-Seeking Strategies.} We first characterized students' overall SRL functions and the epistemic function of the assistance they sought using the validated coding procedure described in~\autoref{sec:codebook-dev}.

\begin{figure}
    \centering
    \subfloat[\centering \textbf{Distribution of Coded SRL Behaviors and the Help-Seeking Content Within REQUEST turns} ]{{\includegraphics[width=6cm]{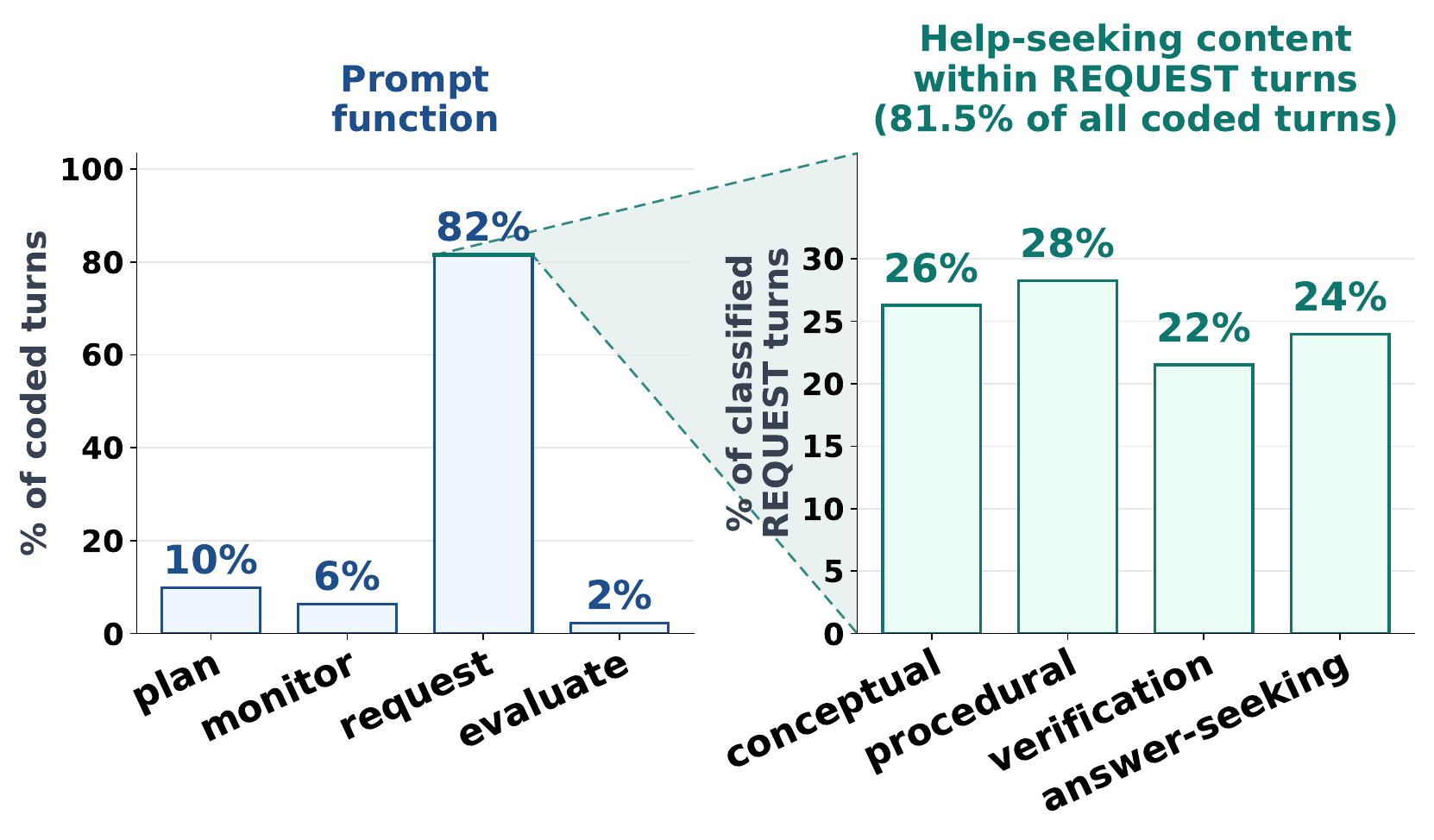} }}%
    \subfloat[\centering \textbf{Distribution of students' subsequent SRL functions following adequate and low-quality tutor responses}]{{ \includegraphics[width=6cm]{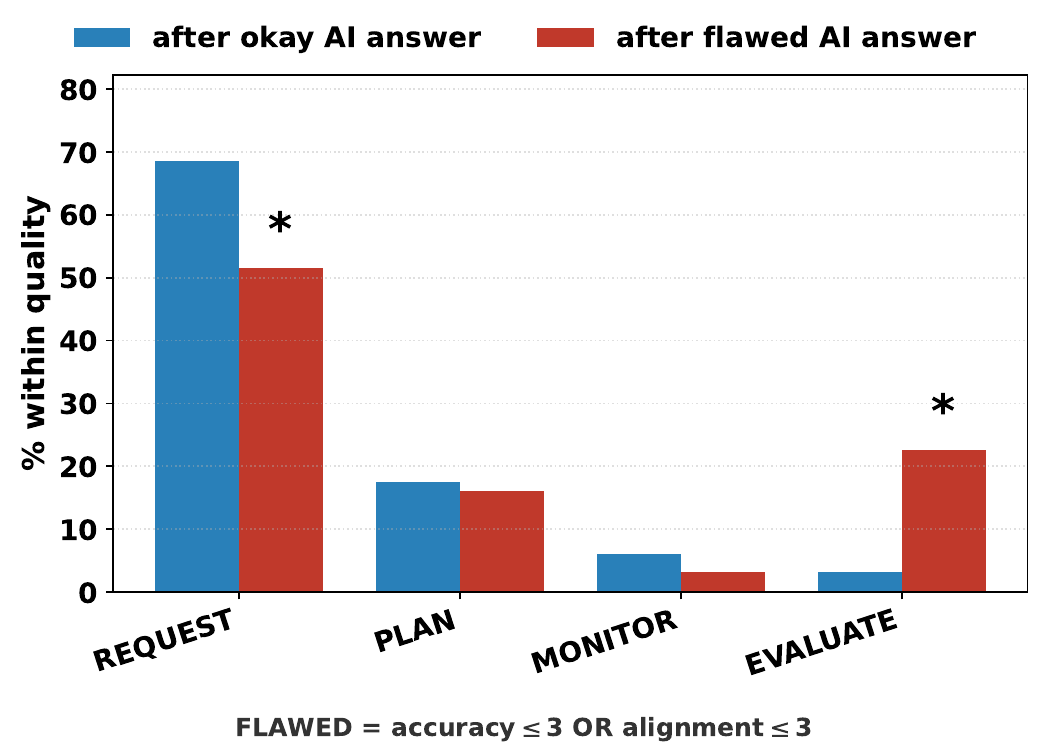} }}%
    \caption{\centering Distributions of students' coded interaction behaviors and behavioral indicators of epistemic vigilance.}%
    \label{fig:chat_descriptions}%
\end{figure}

As shown in sub-figure (a) of~\autoref{fig:chat_descriptions}, interactions were strongly dominated by requests for assistance. Of the 880 substantive turns, 717 (81.5\%) were classified as \textit{REQUEST}, compared with 9.9\% \textit{PLAN}, 6.4\% \textit{MONITOR}, and 2.2\% \textit{EVALUATE}. Thus, although requesting assistance is itself part of self-regulation, other parts of the cycle were comparatively rare in the dialogue. Students thus rarely communicated their state of knowledge (\textit{MONITOR}) or articulated how they intended to advance their learning (\textit{PLAN}) before seeking assistance. This provides limited observable evidence that students themselves were framing their epistemic goals and trajectories, potentially leaving greater space for the AI to do so.

The epistemic function of these requests was more evenly distributed. Among the 717 \textit{REQUEST} turns, 28\% were procedural, 26\% conceptual, 24\% answer-seeking, and 22\% verification/ validation-seeking. Understanding-oriented requests---conceptual and procedural assistance intended to support continued understanding or problem solving---therefore accounted for 54\% of Help-seeking, compared with 46\% answer-seeking or validation. Students thus relied extensively on the tutor for assistance, while differing in whether that assistance primarily supported further knowledge construction or sought an answer or confirmation.

Taken together, these results indicate limited observable epistemic proactivity in students' overall interaction profiles, despite substantial behavioral engagement and important variation in the epistemic function of their help-seeking behaviors. Indeed, we have little explicit evidence that they were themselves initiating, monitoring, or evaluating the broader epistemic trajectory, and nearly half of their help-seeking strategies closure-oriented requests.

\paragraph{Epistemic Vigilance.}
We next examined students' reactions to low-quality tutor responses as an exploratory indicator of expressed epistemic vigilance. A tutor response was classified as low quality when either its accuracy or its alignment with the student's preceding epistemic need was rated $\leq3$, following~\autoref{sec:codebook-dev}. Only 3.9\% of eligible responses met this criterion, producing a strong ceiling in tutor-response quality.

Students' subsequent SRL behavior differed according to response quality, $\chi^2(3)=31.56$, $p=.0006$. Most notably, explicit \textit{EVALUATE} turns increased from 3.1\% following adequate responses to 22.6\% following low-quality responses (adjusted residual $z=5.50$); see sub-figure (b) in~\autoref{fig:chat_descriptions}. Among subsequent turns classified as \textit{REQUEST}, however, response quality was not associated with change in the help-seeking strategy, $\chi^2(3)=1.59$, $p=.68$.

These results suggest that school children were somewhat sensitive to low-quality answers received by AI. They do not, however, establish that students consistently detected problematic assistance, nor that the response itself caused the subsequent behavior. Because low-quality responses were rare, and students' behavior was dominated  by requests, opportunities to observe expressed epistemic vigilance and reactions to it were limited; this analysis was therefore treated as exploratory and Epistemic Vigilance was not included as a student-level predictor in subsequent outcome models.

\subsection{RQ2: Within-Session Dynamics of Epistemic Proactivity}
\begin{figure}%
    \centering
    \subfloat[\centering \textbf{Significant changes in SRL Functions and Help-Seeking strategies across Normalized Session Position.} ]{{\includegraphics[width=6cm]{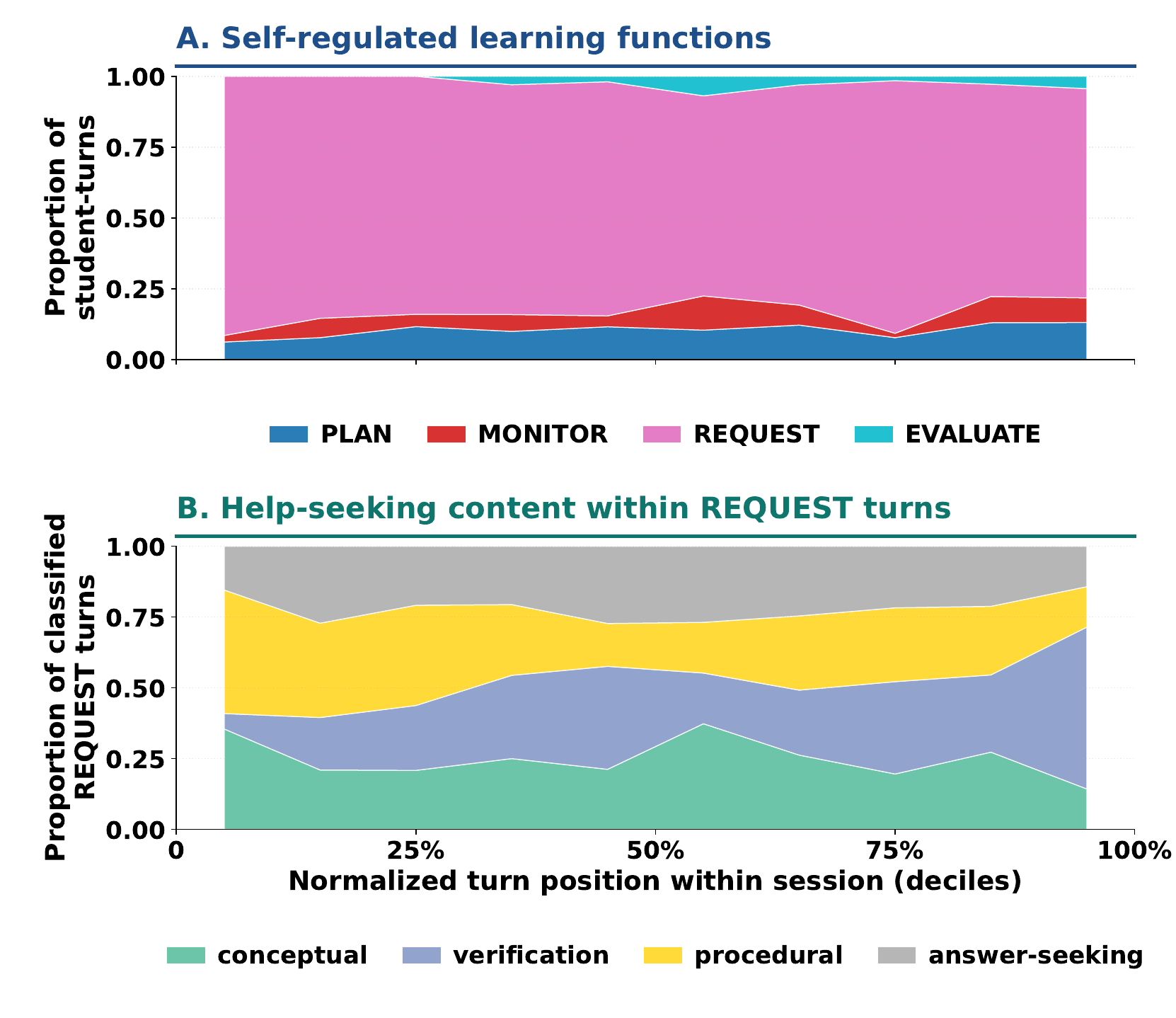} }}%
    \subfloat[\centering \textbf{Distribution of SRL and Help-Seeking Temporal Change Indices}]{{ \includegraphics[width=6cm]{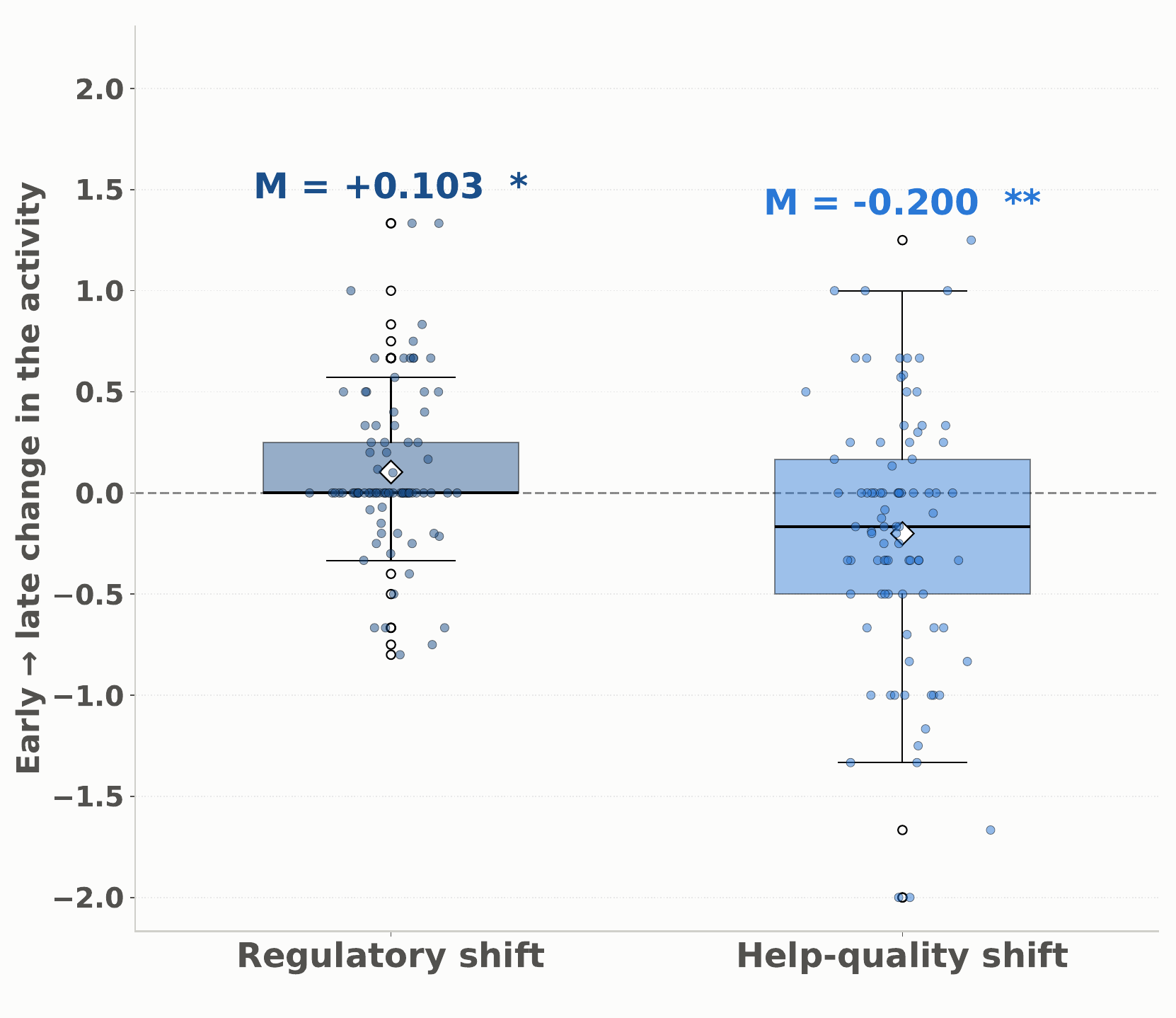} }}%
    \caption{\centering Within-session development of students' SRL functions and Help-Seeking strategies.}%
    \label{fig:temp_descriptions}%
\end{figure}

\paragraph{Directional change across the session.} As discussed in~\autoref{sec:features-def}, we first examined whether students' overall SRL functions and help-seeking strategies changed as the interaction unfolded. At the cohort level, the composition of both dimensions varied across normalized session position: SRL, $\chi^2(27)=43.50$, $p=.023$, and Help-Seeking, $\chi^2(27)=61.60$, $p<.001$; see sub-figure (a) of~\autoref{fig:temp_descriptions}. These differences indicate temporal variation in the relative presence of the four categories, but not its direction.

To characterize directional change, we used the student-level trajectory indices defined in~\autoref{sec:features-def}. Positive SRL values indicate an early-to-late shift from \textit{REQUEST} toward greater explicit \textit{PLAN}, \textit{MONITOR}, and \textit{EVALUATE} activity, whereas positive help-seeking values indicate a shift from answer- and verification/validation-seeking toward more understanding-oriented conceptual and procedural requests. A trajectory index was calculated only when the relevant behavioral dimension was observable in both halves of the interaction; otherwise the score was left undefined.

SRL showed a modest average shift toward greater explicit regulatory activity; see sub-figure (b) of~\autoref{fig:temp_descriptions}. In contrast, the average Help-Seeking trajectory was negative ($M=-.20$), corresponding to a decline in the relative share of understanding-oriented requests from the first to the second half of the interaction.  The two trajectories showed essentially no linear association ($r=.02$, $p=.88$), indicating that changes in explicit SRL activity were not systematically coupled with changes in the epistemic function of students' Help-Seeking and that different components of epistemic proactivity can develop independently.

\paragraph{Local sequential persistence.} Session-level trajectories capture broader directional change but not how behaviors are organized locally. Similar overall trajectories may arise from very different local dynamics: students may alternate between strategies, or repeatedly remain within the same mode of interaction. We therefore complemented them with first-order transition analyses examining whether particular behaviors were more likely than expected to persist or lead to others.

\begin{figure}
\centering
\includegraphics[width=.7\textwidth, clip]{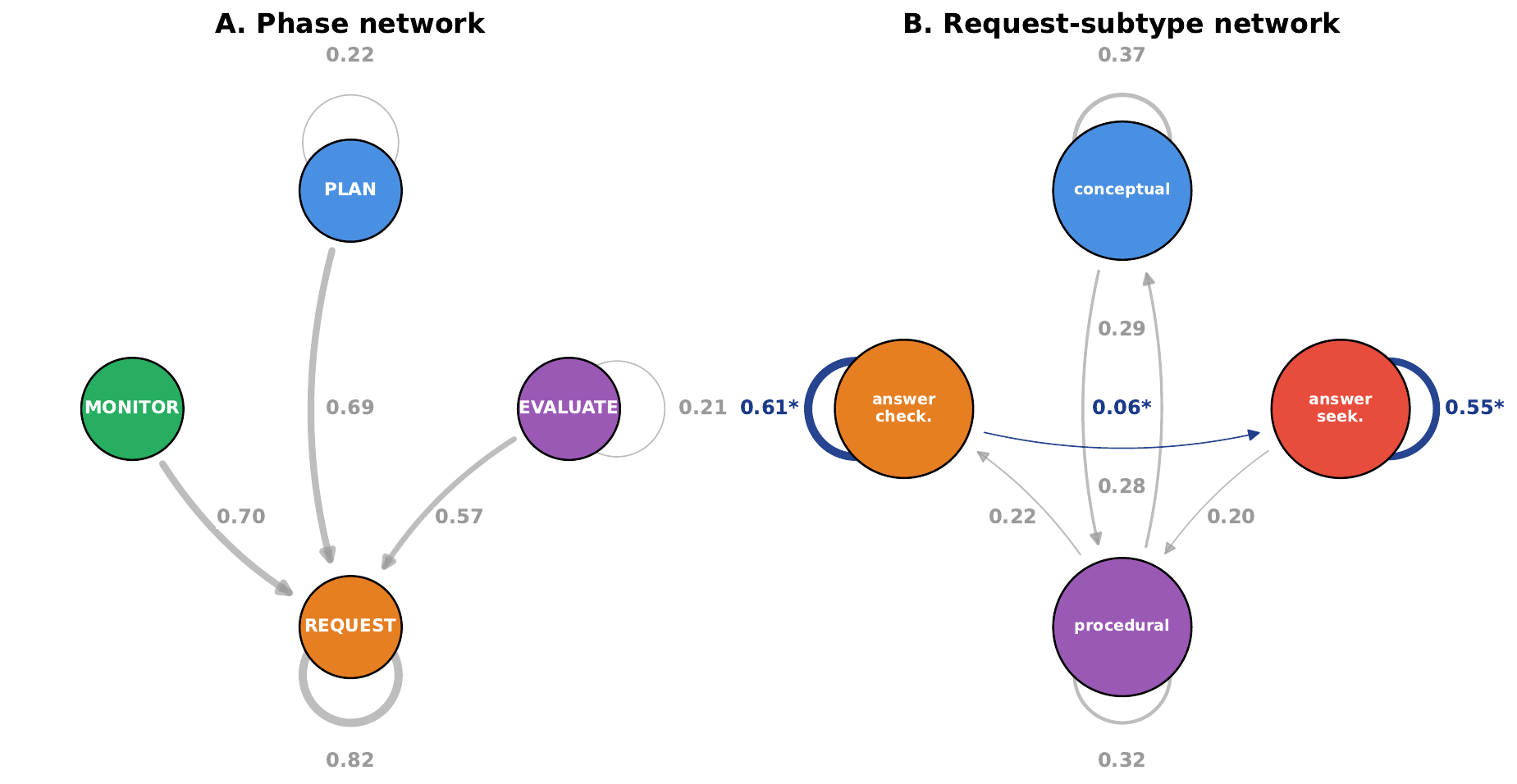}
\caption{Local transition networks for students' SRL functions and Help-Seeking strategies.} 
\label{fig:tna}
\end{figure}

No SRL transition occurred more frequently than expected under the within-student permutation null after FDR correction (smallest $p_{\mathrm{FDR}}=.45$; see~\autoref{fig:tna}). Thus, although transitions involving \textit{REQUEST} were descriptively frequent, none showed above-chance local persistence.

For Help-Seeking, however, showed a different pattern, especially for closure-oriented requests: verification/validation-seeking was followed by another verification/validation-seeking request with probability $P=.61$, 95\% CI $[.46,.73]$, and answer-seeking by another answer-seeking request with probability $P=.55$, 95\% CI $[.39,.66]$. Both self-transitions exceeded the within-student permutation null ($p_{\mathrm{FDR}}=.008$), whereas conceptual and procedural self-transitions did not ($p_{\mathrm{FDR}}=.18$ and $.63$, respectively). Thus, verification- and answer-seeking showed local persistence beyond what would be expected from their overall presence in the conversation: once a request took either form, the following request was more likely to remain in the same help-seeking mode.

\subsection{RQ3: Static and Dynamic Interaction Features and Links with subsequent Performance without AI}
Finally, we examined whether students' AI-interaction features were associated with subsequent post-test performance beyond pre-test performance (i.e. prior knowledge estimation), and whether dynamic features provided information beyond static, whole-session summaries. Of the 95 students with complete pre- and post-test data, six did not have a computable Help-Seeking trajectory because they had no subtype \textit{REQUEST} in one half of the interaction. All models in this section were therefore estimated on the same matched sample of 89 students.

\paragraph{Static interaction features.}
The static block comprised four static whole-session features: the proportion of epistemically-proactive SRL turns, i.e. classified as \textit{PLAN}, \textit{MONITOR}, or \textit{EVALUATE}; the proportion of help-seeking proactive requests, classified as conceptual or procedural; total chat time; and mean AI-response review time. We add the two latter as previous work has used them as proxies for behavioral engagement in AI environments~\cite{liu2026behavioral}. Together, these captured students' overall regulatory and help-seeking composition as well as interaction amount and pace.

Our results show that pre-test performance alone accounted for $R^2=.161$ of post-test variance. Adding the static features increased explained variance only to $R^2=.184$, $\Delta R^2=.023$, $F(4,83)=0.57$, $p=.684$; see~\autoref{tab:regression-results}. Thus, students' whole-session interaction composition, chat duration, and AI-response review time did not account for detectable additional variance in the subsequent performance beyond prior knowledge, once AI assistance was removed. These findings therefore caution against treating the overall epistemic function of students' requests, greater time spent interacting with AI, or longer exposure to its responses as straightforward indicators of more productive learning or better subsequent performance without AI.

\begin{table}[t]
\centering
\caption{Hierarchical regression predicting subsequent post-test performance without AI ($n=89$).}
\label{tab:regression-results}
\small
\begin{tabular}{lccc}
\toprule
 & Step 1 & Step 2 & Step 3 \\
Predictor & $\beta$ & $\beta$ & $\beta$ \\
\midrule
Pre-test performance
    & .402 & .416 & .415 \\

\addlinespace
\multicolumn{4}{l}{\textbf{Static interaction features}}\\
Share PLAN/MONITOR/EVALUATE
    & -- & -.09 & -.09 \\
Share conceptual/procedural HS
    & -- & -.31 & -.31 \\
Total chat time
    & -- & .13 & .13 \\
Mean AI-answer review time
    & -- & -.04 & -.04 \\

\addlinespace
\multicolumn{4}{l}{\textbf{Dynamic interaction features}}\\
SRL trajectory
    & -- & -- & .09 \\
Help-seeking trajectory
    & -- & -- & .34*** \\
Answer-seeking self-transition
    & -- & -- & -.25 \\
Verification-seeking self-transition
    & -- & -- & -.12 \\

\midrule
$R^2$
    & .161 & .184 & .318 \\
Adjusted $R^2$
    & .151 & .135 & .241 \\
$\Delta R^2$
    & -- & .023 & .134 \\
$F$
    & -- & $F(4,83)=.57$ & $F(4,79)=3.89$ \\
$p$
    & -- & .684 & $<$.001 \\
\bottomrule
\end{tabular}

\vspace{1mm}
\footnotesize
\textit{Note.} Standardized regression coefficients are reported. \\
Step 1 included pre-test performance; Step 2 added four static interaction
features; \\Step 3 added four dynamic interaction features.
*** $p<.001$.
\end{table}

\paragraph{Dynamic interaction features.}
The dynamic block comprised the SRL and Help-seeking trajectory indices together with the answer-seeking and verification/validation-seeking self-transition measures identified in the section above. Adding these features to pre-test performance accounted for an additional $\Delta R^2=.112$, $F(4,84)=3.21$, $p=.017$. More importantly, after the static block was already included, the dynamic features increased explained variance from $R^2=.184$ to $R^2=.318$, $\Delta R^2=.134$, $F(4,79)=3.89$, $p=.006$; see~\autoref{tab:regression-results}. The reverse comparison did not show a corresponding contribution of the static features: after pre-test performance and the dynamic block were entered, adding the static features contributed $\Delta R^2=.045$, $F(4,79)=1.29$, $p=.281$. These findings thus suggest that temporal features contained information about the subsequent performance without AI that was not captured by students' aggregate interaction patterns.

This incremental association was primarily driven by the development of students' Help-Seeking. In the full model, the Help-seeking trajectory was positively associated with the post-test performance when AI assistance was removed, $\beta=.34$, 95\% CI $[.15,.53]$, $p<.001$ (adjusted $p=.002$), uniquely accounting for $\Delta R^2=.110$. Students whose requests became relatively more understanding-oriented---toward conceptual and procedural assistance rather than epistemic closure-seeking---subsequently performed better without AI assistance. It is important to note here, however, that the cohort-level showed an average shift in the opposite direction.

No comparable association was detected for the SRL trajectory, $\beta=.09$, 95\% CI $[-.10,.28]$, $p=.347$, answer-seeking self-transition, $\beta=-.25$, 95\% CI $[-.54,.04]$, $p=.090$, or verification-seeking self-transition, $\beta=-.12$, 95\% CI $[-.40,.17]$, $p=.421$. These SRL null estimate should, however, be interpreted cautiously because it was strongly concentrated at zero.

Taken together, the results suggest that the direction in which students' help-seeking developed across the session carried information about the subsequent performance without AI that was not captured by aggregate interaction characteristics or local sequential persistence.

Taken together, our results show that despite the cohort's overall limited epistemic proactivity profile and average drift toward closure-oriented help-seeking, students who developed in the opposite direction performed better subsequently without AI.

\section{Discussion and Practical Implications}
This study examined how Grade-9 school students managed epistemic proactivity when using a general-purpose LLM for open-ended mathematical-modeling practice. We conceptualized this through the SRL framework, complemented by help-seeking and epistemic vigilance, and examined how these behaviors were expressed in students’ conversations with the LLM. Three findings define the overall picture. First, students showed limited observable epistemic proactivity overall: their interactions were strongly request-dominated, with relatively little explicit planning, monitoring, or evaluation, although they were more likely to evaluate AI responses when those responses were of low quality. Second, within-chat change in epistemic proactivity behaviors was uneven and informative: help-seeking became less understanding-oriented on average while explicit SRL activity increased modestly, so different aspects of epistemic proactivity moved in different directions. Finally, individual differences in the help-seeking trajectory mattered, with students whose requests became relatively more understanding-oriented subsequently performing better without AI.

\subsection{RQ1: Students Showed Limited Epistemic Proactivity Profiles Overall.}
Students relied extensively on the AI for assistance. Although slightly more than half of their requests sought conceptual or procedural support, interactions were dominated by requests, with relatively few explicit instances of other SRL moves such as planning, monitoring, or evaluation. This pattern should not be interpreted as complete epistemic reactivity: requesting assistance is itself part of regulation, and regulatory processes that remain internal cannot be observed from conversational traces. Nevertheless, the open-ended task required students themselves to identify what they needed to understand and determine how AI should contribute. The scarcity of explicit regulation therefore suggests that students often made limited observable contributions to structuring the broader learning trajectory. This is indeed be particularly challenging for adolescent learners whose information-search and self-regulatory capacities are still developing~\cite{ruggeri2016sources}, especially given the additional regulatory demands that open-ended AI environments may impose~\cite{tankelevitch2024metacognitive,ruggeri2016sources}.

The epistemic-vigilance results provided a slightly different picture. Students became more explicitly evaluative when AI assistance was inaccurate or poorly aligned, suggesting that they were not uniformly uncritical users. This contrasts with evidence that learners can overlook or adopt erroneous AI-generated information~\cite{noster2024pre,bulut2026ai,abdelghani2026illusion}. However, such opportunities were rare in our data because most AI responses were rated as accurate and well aligned, limiting what can be inferred about students' vigilance more generally. In this context, it is also important to consider that highly fluent and generally adequate assistance may itself create a regulatory challenge: it provides few cues that careful evaluation is required, even though accurate responses may still be poorly calibrated to learners' knowledge, insufficiently explanatory, or pedagogically suboptimal~\cite{pepin2025scoping,Joshi2025ELI-Why:Explanations}. Future work should therefore examine whether prolonged exposure to mostly reliable AI increases perceived source reliability and, in turn, gradually reduces evaluative effort~\cite{kidd2023}.

\subsection{RQ2: SRL and Help-Seeking Strategies Showed Diverging Development Across the Session.} Students' epistemic proactivity was not stable and did not develop uniformly. Their SRL behavior shifted modestly away from request-dominated interaction toward more explicit regulation, while their help-seeking became less understanding-oriented and more focused on answers and verification (epistemic closure). Although these trajectories may appear contradictory, they capture different aspects of epistemic responsibility. Students could become more explicit in monitoring or regulating the interaction while simultaneously asking the AI to perform more of the underlying knowledge-building work. One possible explanation is that sustaining both explicit regulation and understanding-oriented help-seeking places substantial cognitive and metacognitive demands on learners~\cite{tankelevitch2024metacognitive}. Answer- and verification-seeking also showed local persistence, suggesting that once students adopted these forms of assistance, they were more likely than expected to repeat them in the following request. This is consistent with the view that over-reliance may emerge progressively through interaction with AI, as users adapt to the ease of delegating cognitive work, rather than appearing as an immediate and fixed mode of use~\cite{gerlich2025ai}.

One possible explanation for the shift in help-seeking is adaptation to the low cost of obtaining AI assistance. Once explanations, answers, and confirmation become immediately available, while the longer-term learning consequences of delegating this work are not immediately visible, relying on the system to carry more of the task may become increasingly attractive~\cite{misiejuk2026cognitive}. Here, a related challenge is that learners must regulate their behavior with respect not only to the immediate task, but also to the knowledge and competencies they are trying to develop beyond it. Making these longer-term learning goals more explicit may therefore help influence how students decide what assistance to seek~\cite{oakley2026memory}. A complementary explanation concerns the fit between assistance and learners' evolving task representations and available knowledge. Conceptual or procedural explanations requested before sufficient relevant knowledge has been activated may be difficult to process or integrate, particularly when a general-purpose LLM is poorly calibrated to the learner's current understanding~\cite{Seufert2018TheLoad,Joshi2025ELI-Why:Explanations}. Students may consequently shift toward answer-seeking or verification as a lower-cost way to continue and manage cognitive demands, whereas movement toward conceptual or procedural assistance may reflect a better match between the support requested and the knowledge available to make use of it.

\subsection{RQ3: Help-Seeking Strategies Development, rather than static AI-use measures, Was Associated with Better Later Performance Without AI.} The clearest association between student--AI interaction features and subsequent performance without AI concerned the broader development of students' help-seeking, rather than either the overall prevalence or local persistence of particular strategies. Students whose requests increasingly shifted toward conceptual and procedural assistance subsequently performed better without AI than students whose help-seeking shifted toward answer- and verification-seeking. Although answer- and verification-seeking showed local persistence across consecutive requests, these self-transitions were not themselves associated with the post-test performance without AI. Thus, occasional use or repetition of particular help-seeking strategies did not distinguish subsequent performance; what did was the broader direction in which help-seeking developed across the interaction. This supports our hypothesis that individual request types should not be treated as intrinsically productive or unproductive independently of where they occur within the learning process~\cite{ruggeri2022introduction}, and points instead to how epistemic responsibility is organized over time.

These results are consistent with broader accounts of cognitive offloading, which suggest that delegating cognitive and metacognitive work can be beneficial or detrimental depending on when it occurs and on the learner's available resources, and extend this principle to AI-supported learning environments~\cite{ruggeri2022introduction,Seufert2018TheLoad,risko2016cognitive}. Understanding-oriented help-seeking is itself cognitively demanding, as are the processes of knowledge construction and self-regulation that students need to carry out during learning. Students may therefore benefit from strategically deciding which parts of the work to offload to the AI and which to retain. Offloading some reasoning may be useful when it frees cognitive resources for activities that are more important at that moment, whereas retaining responsibility may be necessary when performing the reasoning itself is essential for building or consolidating knowledge. Productive AI use may thus depend less on minimizing or maximizing delegation than on regulating when delegation is helpful and when learners need to carry out the cognitive work themselves.

Our findings also challenge the idea that successful learning with AI can be inferred primarily from the quality of the information students receive. They instead highlight the importance of the epistemic work involved in seeking and using that assistance. Indeed, in our data, AI-response quality did not significantly differ according to whether the preceding request was understanding- or closure-oriented: students generally received high-quality assistance irrespective of the epistemic orientation of their request. Yet the development of their help-seeking was associated with their subsequent performance without AI. Thus, receiving high-quality AI-generated information does not necessarily imply that learners have engaged in the cognitive activity required to learn from it. This extends to AI-supported learning broader concerns that exposure to relevant information---and even its evaluation---does not guarantee that it will subsequently be accessible without assistance~\cite{yan2026agentivism,Yan2025DistinguishingAI}. Durable learning depends on learners engaging in the operations through which knowledge is actively sought and retrieved~\cite{oakley2026memory,ruggeri2022introduction}. This interpretation is consistent with evidence that learners benefit when they generate or structure their own work before consulting AI~\cite{wong2026think}, and that unrestricted solution-giving can improve assisted performance without producing comparable subsequent performance without AI~\cite{bastani2025generative}.

Taken together, the results suggest that limited epistemic proactivity at the aggregate level does not mean that its temporal organization is irrelevant to learning. Rather, how learners redistribute epistemic responsibility over the interaction---particularly through their help-seeking strategies---can carry information about what they are subsequently able to do without AI. Productive AI-supported learning may therefore depend less on particular prompt types than on whether learners retain or reclaim responsibility for cognitive and regulatory work that is itself important for learning.

\paragraph{Practical implications.} For educational AI design, these findings suggest moving beyond evaluating isolated prompts toward conversation-sensitive analysis and scaffolding. Individual answer- or verification-seeking requests should not necessarily be treated as problematic; rather, systems could monitor whether help-seeking progressively becomes more closure-oriented or whether learners contribute less to advancing their own understanding over time, which may indicate increasing offloading. More fundamentally, AI assistance should be calibrated so that the system does not perform the very cognitive operations that learners are expected to develop. If the learning objective requires students to retrieve and apply knowledge, explain a concept, connect ideas, or justify a solution, the AI should support rather than replace these activities. This could involve first eliciting learners' current understanding and knowledge gaps, prompting an initial attempt, providing hints or partial information, or asking learners to retrieve and connect relevant concepts before more complete assistance is given. AI systems could also maintain an estimate of learners' current knowledge and use it to calibrate the assistance they provide. When a learner has not yet acquired a target knowledge component, the system could avoid immediately taking over the reasoning required to practice it and instead scaffold the learner's own attempt. Conversely, once that component is sufficiently established, delegating it to the AI may be appropriate as doing so can free resources for a more advanced learning objective. The objective is therefore not to minimize AI use or prevent offloading, but to preserve learning-relevant cognitive work while allowing other demands to be strategically delegated. This follows the assistance-dilemma principle~\cite{koedinger2007exploring} and is consistent with evidence that pedagogically constrained LLM support can better preserve subsequent performance when AI assistance is removed, than unrestricted solution provision~\cite{bastani2025generative}.

For teachers, this means that AI literacy should extend beyond teaching students how to formulate effective prompts or operate AI tools. Before consulting AI, teachers should encourage students to articulate what they are trying to learn, what they already understand, what remains unclear, and, where possible, make a first attempt themselves. This can help establish a sequence in which AI augments students' own reasoning---for example, by critiquing, extending, or helping them revise an initial attempt---rather than becoming the default go-to whenever they encounter difficulty. More broadly, epistemic proactivity with AI can itself become an instructional objective. Students need to learn not only \emph{how} to obtain useful assistance, but also \emph{which} cognitive work can appropriately be delegated and which should remain with them because performing that work is itself part of what they are learning. This requires helping students distinguish longer-term learning goals from short-term outcomes such as immediate task completion, and to regulate their AI use accordingly. In this sense, teaching students how to formulate questions that require them to articulate knowledge gaps, retrieve relevant information, or seek deeper understanding may also contribute to more productive use of the system. Students also need explicit practice evaluating AI responses even when they appear plausible and correct, for example by identifying remaining uncertainties, justifying whether an explanation is sufficient, or checking it against their own knowledge or other sources. At the same time, the importance of foundational instruction should not be overlooked: prior knowledge remained the strongest predictor of subsequent performance without AI in our study. Building these foundations therefore remains essential, and AI should not routinely perform the very retrieval, reasoning, or application processes through which students are expected to acquire them. Finally, when having students work with AI, teachers should avoid treating isolated answer- or verification-seeking requests as inherently unproductive; greater attention may instead be warranted when these strategies become increasingly dominant over the course of an interaction without a return to more understanding-oriented activity.

\section{Limitations and Future Work}
Several limitations qualify these findings and point toward future work. First, the interaction took place within a single, relatively short learning session. The trajectories identified here therefore capture within-session changes in students' observable behavior rather than the development of stable SRL or Help-Seeking skills. Repeated interactions with AI may produce different dynamics as learners gain experience with the system, calibrate their trust, and develop more stable strategies for regulating AI assistance. Longitudinal studies are therefore needed to examine whether the trajectories observed here persist, intensify, or change across repeated use, and whether students can learn to maintain epistemically proactive patterns over time. Relatedly, the post-test immediately followed the interaction, so the present findings concern subsequent performance rather than long-term retention.

Second, our analysis focused primarily on domain-general regulatory and help-seeking processes. We initially sought to complement these measures with students' progression through mathematical modeling processes, but the available modeling framework could not be applied with sufficient reliability to these open-ended dialogues and was therefore excluded from the analysis. This limits our understanding of how changes in epistemic proactivity were intertwined with students' mathematical reasoning---for example, whether understanding-oriented help-seeking accompanied productive movement between important modeling steps or helped students overcome particular difficulties. Future work should develop more reliable, domain-grounded approaches for tracing learners' knowledge states within open-ended AI conversations and examine how these interact with SRL and help-seeking over time.

Third, the associations between interaction dynamics and subsequent performance are observational and should not be interpreted causally. In particular, the present data cannot establish whether increasingly answer-oriented help-seeking impairs learning, whether students who struggle are more likely to shift toward such strategies, or whether both reflect other learner characteristics. Experimental work manipulating how and when AI systems provide assistance will be necessary to test the mechanisms suggested by the present results. Our measures also capture only regulation that becomes visible in students' dialogue; planning, monitoring, or evaluation may occur internally without being verbalized. Similarly, the epistemic-vigilance analysis should remain exploratory because low-quality AI responses were rare, limiting our ability to characterize how systematically students evaluated AI output.

Finally, the study involved Grade-9 students completing mathematical modeling activities with one general-purpose LLM. The extent to which the observed trajectories generalize to other age groups, domains, tasks, models, or more pedagogically constrained AI tutors therefore remains an open question. In addition, participants' gender were not collected sought to avoid collecting potentially sensitive personal information from minors, preventing us from examining whether patterns of epistemic proactivity or their associations with performance without AI differed across sex or gender groups. This further limits the generalizability of the findings. Future studies should examine these dimensions explicitly, alongside replication across different learner populations and instructional contexts.

\section{Conclusion}
Productive learning with AI may depend not only on what learners ask, but on how their relationship with AI assistance develops over time. In this study, grade-9 students showed limited observable epistemic proactivity overall when learning with a general-purpose LLM, yet individual differences in how their help-seeking developed were meaningfully associated with later performance without AI. Although help-seeking became less understanding-oriented on average, students who developed in the opposite direction performed better subsequently, whereas static interaction patterns and local persistence did not provide comparable information.

These findings position epistemic proactivity as a dynamic property of AI-supported learning rather than a characteristic that can be fully understood from isolated prompts or session-level averages. Supporting productive AI use may therefore require moving beyond teaching students how to formulate good prompts toward helping them sustain understanding-oriented patterns of interaction over time.

\bibliographystyle{unsrt}  
\bibliography{cas-refs-curated}  %%% Remove comment to use the external .bib file (using bibtex).


\end{document}